\documentclass[manuscript,graphicx]{aastex}

\begin{document}

\voffset = 0.0truecm

\newcommand{\gray}{$\gamma$-ray\  } \newcommand{\grays}{$\gamma$-rays\
} \newcommand{\etal}{{\it  et al.\ }}  \newcommand{\lya}{Ly$\alpha$\ }
\newcommand{\epr}{e-print  astro-ph/  }  \newcommand{\e}  {\epsilon  }
\newcommand{\mic}{$\mu$m\  }  \slugcomment{Astrophysical
Journal Letters 652, L9 (2006)}
\title{A  Simple Analytic  Treatment of  the  Intergalactic Absorption
Effect in Blazar Gamma-Ray Spectra}

\author{F.W.    Stecker}  \affil{NASA/Goddard  Space   Flight  Center}
\authoraddr{Greenbelt, MD 20771} \email{stecker@milkyway.gsfc.nasa.gov}
\author{S.T.   Scully}  \affil{Department  of Physics,  James  Madison
University}         \authoraddr{Harrisonburg,         VA        22807}
\email{scullyst@jmu.edu}

\begin{abstract}

We derive a new and user friendly simple analytic approximation for  
determining the
effect of intergalactic absorption on the \gray spectra of TeV blazars
the energy range $0.2 ~{\rm TeV} < E_{\gamma} < 2$ TeV and the redshift
range $0.05 <  z < 0.4$.   In these  ranges, the form  of the
absorption    coefficient    $\tau(E_{\gamma})$    is    approximately
logarithmic.  The  effect of  this  energy  dependence  is to  steepen
intrinsic  source  spectra such  that  a  source  with an  approximate
power-law  spectral index  $\Gamma_{s}$ is  converted to  one  with an
observed spectral index $\Gamma_{o} \simeq \Gamma_{s} + \Delta \Gamma (z)$
where $\Delta  \Gamma (z)$  is a linear function  of $z$ in the  
redshift range 0.05 --  0.4.  We apply this approximation to the 
spectra of 7 TeV blazars.

\end{abstract}


\section{Introduction}

Stecker, Malkan \& Scully (2006)  (SMS) have used recent {\it Spitzer}
observations (Le  Floc'h \etal 2005, Perez-Gonzalez  \etal 2005) along
with other data on  galaxy luminosity functions and redshift evolution
in order  to make  a detailed evaluation  of the  intergalactic photon
density as a function of both energy  and redshift for $0 < z < 6$ for
photon energies from .003 eV to the Lyman limit cutoff at 13.6 eV in a
$\Lambda$CDM universe with $\Omega_{\Lambda}  = 0.7$ and $\Omega_{m} =
0.3$. They  then used their calculated  intergalactic photon densities
to calculate  the optical  depth of the  universe, $\tau$,  for \grays
having energies from 4 GeV to  100 TeV emitted by sources at redshifts
from  ~0  to  5.  They  also  gave  a  parametric fit  with  numerical
coefficients for $\tau(E_{\gamma},z)$.

As  an   example  of  the   application  of  the   detailed  numerical
determination  of  the optical  depth,  they  calculated the  absorbed
spectrum of  the blazar PKS  2155-304 at $z  = 0.117$ and  compared it
with  the  spectrum observed  by  the  H.E.S.S.   air Cherenkov  \gray
telescope  array.  It  was  noted  that a  steepening  in an  $E^{-2}$
power-law differential  photon source spectrum for  PKS 2155-304 would
be  steepened by  approximately  one  power to  $E^{-3}$  as would  be
produced by a \gray opacity with a logarithmic energy dependence.

The purpose of this letter is to generalize this result by determining
fits for approximating  $\tau(E_{\gamma}, z)$ by logarithmic functions
in  order to predict  power-law steepenings  in assumed  blazar source
spectra of the form $E^{-\Gamma_{s}}$  to observed spectra of the form
$E^{-\Gamma_{o}}$ where $\Gamma_{o} \simeq \Gamma_{s} + \Delta \Gamma$
and $\Delta \Gamma$ is determined to be a linear function of $z$.

\section{Calculation}

In order to derive our results, we fit the results from SMS  for 
$\tau(E_{\gamma}, z)$ to a form which
is assumed to be logarithmic in $E_{\gamma}$ in the energy range $0.2~
\rm TeV < E_{\gamma} < 2 ~\rm  TeV $ and which has a linear dependence
on $z$ over the range $0.05 < z < 0.4$. It is important to note that
our linear fit to the $z$ dependence is both {\it qualitatively} and
{\it quantitatively} different from
the linear dependence on redshift which would be obtained for small
redshifts $z << 1$ and which simply comes from the fact that for small
$z$ where luminosity evolution is unimportant and where $\tau \propto
d$, with the distance $d \simeq cz/H_{0} \propto z$, our quantitaitve
fit for the higher redshift range $0.05 < z < 0.4$ comes from the more
complex calculations based on the models of SMS. For this reason,
our linear fits are not simply proportional to redshift.

We choose the form of our fitting function inn both redshift and
energy to be given by

$$\tau(E_{\gamma},z)  = (A  + Bz)  + (C  + Dz)\ln  [E_{\gamma}/(\rm 1~
TeV)],$$

where $A, B, C and D$ are constants.

\begin{deluxetable}{ccccc}
\tabletypesize{\small}
\tablecaption{Optical  Depth Parameters}  \tablewidth{0pt} \tablehead{
\colhead{Evolution Model} & \colhead{A}  & \colhead{B} & \colhead{C} &
\colhead{D} }  \startdata Fast Evolution &  -0.475 & 21.6  & -0.0972 &
10.6 \\ Baseline & -0.346 & 16.3 & -0.0675 & 7.99 \\ \enddata
\end{deluxetable}

If we then postulate an intrinsic source spectrum which can be approximated 
by a power law over this limited energy range of one decade,

$$\Phi_{s}(E_{\gamma})~ \simeq~ KE_{\gamma}^{-\Gamma_{s}},$$

\noindent the spectrum  which will be observed at  the Earth following
intergalactic absorption will be of the power-law form

$$\Phi_{o}(E_{\gamma})~ = ~ Ke^{-(A+Bz)}E_{\gamma}^{-(\Gamma_{s}+C+Dz)}.$$

This can be  compared with the empirically observed  spectrum which is
usually presented in the literature  to be of the power-law form.  The
observed spectral index, $\Gamma_{o}$, will then be given by

$$\Gamma_{o} = \Gamma_{s} + \Delta \Gamma (z) $$

\noindent  where  the  intrinsic  spectral  index  of  the  source  is
steepened by $ \Delta \Gamma (z)~ =~ C + Dz$.

\begin{figure}[h]
\epsscale{.80} \plotone{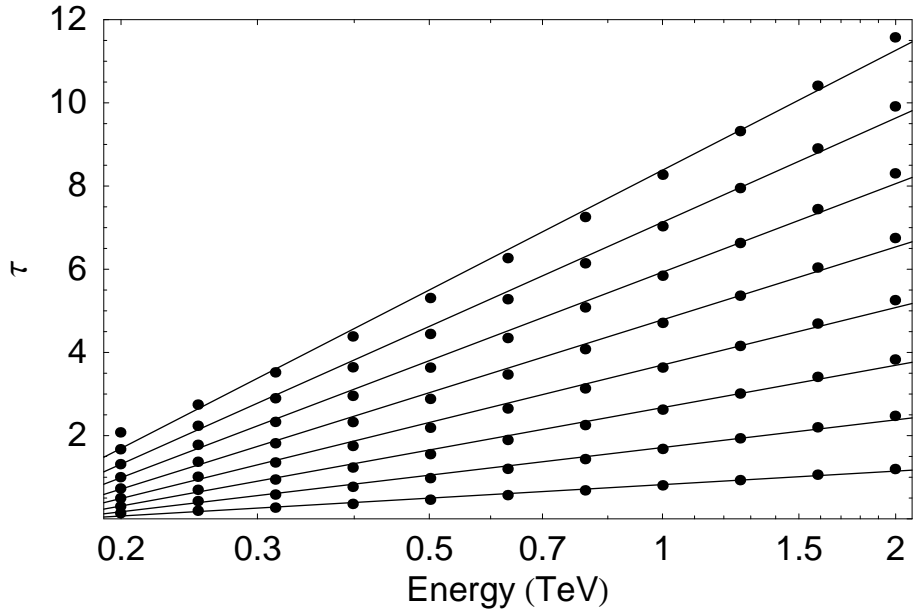}
\caption{  Linear  fits  in  $ln  E_{\gamma}$ to  the  optical  depths
calculated in SMS (points shown) for their fast evolution (FE) case in
the energy range 0.2 TeV to 2.0 TeV for redshifts of (bottom to top)
0.05, 0.10, 0.15, 0.20, 0.25, 0.30, 0.35, and 0.40. A similar result is 
obtained for the SMS baseline (B) case.\label{f1}}

\end{figure}

\begin{figure}[h]
\epsscale{.80}
\plotone{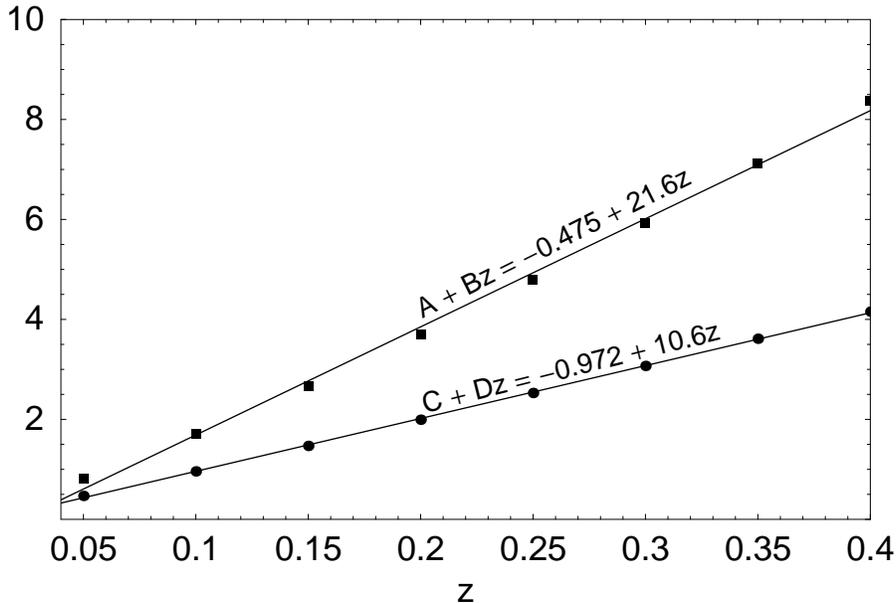}
\caption{ The fits obtained for the linear functions $A + Bz$ and $C + Dz$
shown for the FE case as descibed in the text. A similar result is
obtained for the B case.\label{f2}}

\end{figure}

The  parameters $A, B,  C$, and  $D$ obtained  by fitting  the optical
depths derived for the fast  evolution (FE) and baseline (B) models of
SMS  are given  in Table  1. Figures  \ref{f1} and  \ref{f2}  show the
excellent fits of these parameters to a linear dependence in both $\ln
E_{\gamma}$ and redshift.

\section{Spectral Indices of Individual Sources}

Table 2 gives a list of blazars which have been detected at TeV energies
and for which spectral  indices ($\Gamma_{o}$) have been measured in the
energy range 0.2 -- 2 TeV.  We also give  the observed  redshifts of 
these  sources and  the intrinsic spectral
indices of the sources ($\Gamma_{s}$) derived from the baseline (B)and
fast evolution (FE) models  using our analytic expressions for $\Delta
\Gamma (z)$. A spectral index of less than 2 indicates that the energy
range of the  observation would be below the Compton  peak energy in the
spectral energy distribution, $E_{\gamma}^2\Phi(E_{\gamma})$ of a
synchrotron self-Compton  (SSC) model (Stecker, De Jager \& Salamon 1996). 
In the case  of the blazar PG 1553+113, the
energy  range of  the observation  is from  0.09 to  0.4 TeV  which is
somewhat  below  the  energy  range  we  have  used  in  deriving  our
approximation.

\begin{deluxetable}{ccccc}
\tabletypesize{\small}   
\tablecaption{Spectral Indices for Seven Observed Blazars in the 0.2 -- 2 TeV 
Energy Range}  
\tablewidth{0pt}  
\tablehead{  \colhead{Source}  &
\colhead{$z$}      &     \colhead{$\Gamma_{o}$\tablenotemark{*}}     &
\colhead{$\Gamma_{s}$ (B)} &  \colhead{$\Gamma_{s}$ (FE)} } 
\startdata

Mkn 180 & 0.045 & 3.3 (a) &  3.0  & 2.9  \\  
PKS 2005-489 & 0.071  & 4.0 (b) & 3.5  & 3.4 \\ 
PKS 2155-304  & 0.117 & 3.3 (c)  & 2.4 &  2.2 \\ 
H 2356-309 & 0.165  & 3.1 (d) & 1.9  & 1.5 \\ 
1ES 1218+30 &  0.182 & 3.0 (e) & 1.6 & 1.2 \\
1ES 1101-232 & 0.186 & 2.9 (f) & 1.5 & 1.0 \\
PG 1553+113 & 0.36\tablenotemark{\dagger} & 4.2 (g) & 1.4 & 0.5 \\ 

\enddata
\tablenotetext{*}{MAGIC and H.E.S.S.references for observed spectral indices 
denoted by letters are (a) Albert \etal (2006b), (b) Aharonian  \etal (2005b), 
(c) Aharonian \etal (2005a), (d) Aharonian \etal (2006a),
(e) Albert \etal (2006a), (f) Aharonian \etal (2006b),
(g) Albert \etal (2006b).}

\tablenotetext{\dagger}{Although Albert \etal (2006b) and Aharonian \etal
(2006c) use assumptions about the source spectrum of this object to place 
upper limits on $z$ of 0.78 and 0.74 respectively, the catalogued redshift
of PG 1553+113 is 0.36 (Hewitt and Burbidge 1989).}

\end{deluxetable}

\section{Conclusions}

We have  derived a new and user friendly simple  analytic approximation 
for determining the
effect of intergalactic absorption on  the spectra of \gray sources in
the energy  range 0.2 TeV $ <  E_{\gamma} <$ 2 TeV and the redshift range
$0.05 < z < 0.4$, where the absorption is primarily from interactions
with intergalactic photons in the optical to near infrared wavelength
range.  In these energy and redshift ranges, the form of the absorption
coefficient $\tau(E_{\gamma})$ is approximately logarithmic. The effect of
this  energy dependence is  to steepen  the intrinsic  source spectrum
such  that  a source  with  an  approximate  power-law spectral  index
$\Gamma_{s}$  is converted  to  one with  an  observed spectral  index
$\Gamma_{o}  \simeq \Gamma_{s} +  \Delta \Gamma$  in the  energy range
0.2 -- 2 TeV,  where $\Delta \Gamma (z)$  is a linear function  of $z$ in
the redshift  range 0.05 -- 0.4.  We  have applied this  approximation to
the  spectra of  seven TeV  blazars.  These  power-law approximations,
both observational  and theoretical, are  only useful over  the limited
energy  range indicated,  {\it viz.}  0.2--2 TeV.   The  actual spectra
should exhibit  some curvature. At higher energies  the spectra should
cut off either because of a  natural upper limit to the source spectra
or, more  likely in the case  of high frequency peaked  BL Lac objects
(HBLs),  because  of the increase  in  optical depth  with
energy owing to interactions with the much more numerous intergalactic
far infrared photons. A full numerical treatment of absorption can be made 
using the exact form of  the optical depth as a function  of redshift 
and energy as calculated in  the detailed work of SMS. Examples
of  more  detailed  treatments  of  effects  of  intergalactic  \gray\
absorption are given for Mkn 501  and Mkn 421 by Konopelko \etal (2003)
and for PKS 2155-304  by SMS. However,  because of the
limited energy range of empirical TeV spectral determinations, and because
of  the  significant   statistical  and  systematic  uncertainties  in
the empirical  determinations  of spectral  indices  of  TeV sources,  our
approximate linear relations  can be quite useful in  making rapid and
simple analyses of TeV source spectra.

\end{document}